\documentclass[twocolumn]{aastex631}

\usepackage{amsmath}
\newcommand{\Mch}{\mathcal{M}}

\begin{document}

\title{The origin of the nano-Hertz stochastic gravitational wave background:\protect\\ the contribution from $z\gtrsim1$ supermassive black-hole binaries}

\author[0000-0001-8426-5732]{Jean J. Somalwar}\email{jsomalwa@caltech.edu}
\affil{Cahill Center for Astronomy and Astrophysics, MC\,249-17 California Institute of Technology, Pasadena CA 91125, USA.}

\author[0000-0002-7252-5485]{Vikram Ravi}
\affil{Cahill Center for Astronomy and Astrophysics, MC\,249-17 California Institute of Technology, Pasadena CA 91125, USA.}

\begin{abstract}
The nano-Hertz gravitational wave background (GWB) is a key probe of supermassive black hole (SMBH) formation and evolution, if the background arises predominantly from binary SMBHs. The amplitude of the GWB, which is typically quantified in terms of the characteristic strain, $A_{\rm 1 yr}$, at a frequency $1\,{\rm yr}^{-1}$, encodes significant astrophysical information about the SMBH binary (SMBHB) population, including the mass and redshift distributions of SMBHBs. Recent results from a number of pulsar timing arrays have identified a common-spectrum noise process that is consistent with a loud GWB signal with amplitude $A_{\rm 1 yr}{\sim}2\times10^{-15}$, which is higher than typical predictions $A_{\rm 1 yr} \lesssim 10^{-15}$. These predictions usually assume theoretically-motivated but highly uncertain prescriptions for SMBH seeding and evolution. In this work, we use a simple, flexible model of SMBH evolution to explore the possible range of GWB amplitudes, given observational constraints. In particular, we focus on the possible contribution to the GWB from high redshift ($z\gtrsim 1$) SMBHBs, for which few robust observational constraints exist. We find that the GWB amplitude may be higher than fiducial predictions by as much as ${\sim}0.5$ dex if much of the SMBH mass density was established by $z\sim1$. Beyond pulsar timing constraints, observations of the high redshift SMBH population from the James Webb Space Telescope and the Laser Interferometer Space Antenna will be key for constraining the contribution of high-$z$ SMBHBs to the GWB.
\end{abstract}

\section{Introduction} \label{sec:intro}

Despite their outsize influence, many open questions surround the formation and evolution of supermassive black holes (SMBHs). Although it is well established that $\gtrsim 10^9\,M_\odot$ SMBHs exist at $z\gtrsim 6$, the mechanism that enables such rapid formation is hotly debated \citep[e.g.,][]{Fan2001AJ....122.2833F, Volonteri2003ApJ...582..559V}. The occupation fraction of SMBHs and the physical processes that allow for SMBH binary (SMBHB) formation and merging within a Hubble time are likewise not well understood \citep[e.g.,][and references therein]{Fan2001AJ....122.2833F,Armitage2002ApJ...567L...9A,Milosavljevic2003AIPC..686..201M,Lodato2006MNRAS.371.1813L, Miller2015ApJ...799...98M}.

Upcoming observations will be key to uncovering the origins and evolution of SMBHs. Pulsar timing arrays (PTAs), in particular, probe the population of merging SMBHs \citep[see][for a review]{Burke2019A&ARv..27....5B}. They are primarily sensitive to nano-Hertz gravitational waves (GWs) produced by SMBHBs at approximately milli-parsec separations, within $\sim10^{6}$\,yr of merging. The loudest SMBHBs could be detected as individual sources and the population of SMBHBs is detectable as a stochastic gravitational wave background \citep[GWB; e.g.,][]{Rajagopal19951995ApJ...446..543R,Phinney2001astro.ph..8028P,Wyithe20032003ApJ...590..691W}. This GWB consists of the integrated emission in the nHz band from the full population of SMBHBs over cosmic time.

The GWB, which is the focus of this work, is a probe of both the local and distant SMBHB populations. The characteristic strain spectrum for the GWB can be expressed in terms of an SMBHB population as \citep[][]{Phinney2001astro.ph..8028P, Sesana2008MNRAS}
\begin{gather}
    h_c^2(f) = \frac{4}{\pi f^2} \int_0^{\infty} dz \int_0^\infty d\mathcal{M} \frac{d^2n}{dz d\mathcal{M}} \frac{1}{1+z} \frac{\pi^{2/3}}{3} \mathcal{M}^{5/3} f_r^{2/3} \nonumber \\
    = A_{\rm 1\,yr}^2 \Big(\frac{f}{1\,{\rm yr}^{-1}}\Big)^{-4/3}.
\end{gather}
Here, $f$ is the observed frequency of the GW signal and $f_r = (1+z)f$ is the rest-frame frequency for redshift $z$. The chirp mass is given by $\mathcal{M} = \mu^{3/5} M^{2/5}$ for reduced mass $\mu$ and total mass $M$. $\frac{d^2n}{dz d\mathcal{M}}$ is the SMBHB number density per unit redshift and chirp mass.

In the absence of non-gravitational hardening (and/or softening effects), the strain amplitude $A_{\rm 1yr}$ encodes all astrophysical information about the SMBHB population. This amplitude thus sets constraints on models of SMBH(B) formation and evolution. Tantalizingly, recent PTA datasets show evidence for a common-spectrum noise process that, although so far uncorrelated between pulsars, potentially represents a GWB signal \citep{Arzoumanian2020ApJ...905L..34A,Goncharov20212021ApJ...917L..19G,Chen20212021MNRAS.508.4970C}. For example, the NANOGrav data indicate $A_{\rm 1\,yr^{-1}}=1.92 \times 10^{-15}$ ($5\%{-}95\%$ quantiles $(1.37{-}2.67) \times 10^{-15}$) \citep[][]{Arzoumanian2020ApJ...905L..34A}. If this is indeed a GWB, this result is surprisingly high: predictions for $A_{\rm 1yr}$ are typically $\lesssim 10^{-15}$ \citep[see, e.g.,][and references therein]{2022MNRAS.509.3488I}. However, these predictions are often based on models that assume a theoretically motivated prescription for SMBH seeding and evolution. Other predictions \citep[e.g.,][]{Sesana2013MNRAS.433L...1S,Ravi2015MNRAS.447.2772R} make use of empirical constraints on the SMBH coalescence rate, with a focus on lower redshifts where observations are more constraining at present. Our observational knowledge of SMBH formation and growth across cosmic time is limited, and simulations are not fully able to reproduce observed properties of the known SMBH population at modest and high redshifts \citep[e.g.,][]{Habouzit20222022MNRAS.511.3751H, Natarajan2021arXiv210313932N}. It is possible that we are underestimating the number of very massive SMBHBs beyond the local universe.

Any mistaken assumptions about SMBH formation and evolution will affect the predicted amplitude of the GWB. We are thus motivated to consider: what GWB amplitude is feasible given the most reliable information we have about the local and high redshift SMBH populations? We are particularly curious about the potential contributions from a so far unmodeled population of massive high-redshift mergers, given the turnover in GW detectability with redshift at fixed chirp mass \citep{Rosado20162016PhRvL.116j1102R}. Indeed, evidence is emerging that high-redshift SMBHs fall above local SMBH-galaxy scaling relations by factors of $\gtrsim0.5$\,dex \cite[e.g.,][]{Agarwal2013MNRAS.432.3438A,2016ApJ...816...37V,Neeleman2021ApJ...911..141N}. 

In this work, we constrain the feasible range of GWB amplitudes without assuming a single model that encapsulates both the SMBH seeding physics and evolution. Instead, we develop a simple SMBH model that is consistent with predictions from $\Lambda$CDM and assumes SMBH-galaxy coevolution. We enforce four constraints: (1) the $\Lambda$CDM predictions for dark-matter halo mergers, and a self-consistent model for galaxy growth (Section~\ref{sec:TNG}); (2) consistency with the local galaxy stellar mass -- SMBH mass relation (Section~\ref{sec:BH_gal}); (3) a loose but extant relationship between galaxy stellar mass and SMBH mass at high redshifts (Section~\ref{sec:BH_gal}); and, (4) consistency with the quasar luminosity function (Section~\ref{sec:QLF}).  With these constraints, we calculate the range of possible GWB amplitudes (Section~\ref{sec:results}) and consider the implications for PTA observations and other probes of high-redshift SMBHs (Section~\ref{sec:imp}). We conclude with a discussion of future prospects in the JWST-LISA era (Section~\ref{sec:imp}).

\section{Methodology} \label{sec:methods}

In this section, we describe our methods of specifying the population of merging SMBHBs and calculating the amplitude of the GWB. We begin by describing the adopted galaxy merger trees and then discuss our prescription for populating galaxies with SMBHs. We detail key observational constraints on SMBH populations, with which we require our simulated populations are consistent. We finally discuss the calculation of the GWB amplitude.

\subsection{The Illustris-TNG Simulation Suite and Merger Trees} \label{sec:TNG}

We adopt the galaxy merger trees from the IllustrisTNG simulation suite, which are a set of cosmological simulations that include magnetohydrodynamics. A detailed description of the simulation suite and included physics is provided by \cite{TNG1,TNG2,TNG3,TNG4,TNG5,TNG6,TNG7}. Most pertinent to this work, the suite is able to reproduce observations of the matter and galaxy power spectrum \citep[][]{TNG1} and include a simple prescription for SMBH seeding and evolution that we will discuss shortly.

We fiducially adopt the TNG100 simulation, which has $2\times1820^3$ resolution elements and a $110.7^3$ Mpc$^3$ box-size. The baryon mass resolution is $1.4\times10^6\,M_\sun$.

The IllustrisTNG simulations include a model for SMBH formation and growth \citep{Weinberger2017MNRAS}, which we briefly summarize here. When a halo has reached a threshold mass $M_{\rm FOF} = 5\times10^{10}\,{\rm M_\odot}$, an SMBH with seed mass $M_{\rm seed} = 8\times10^5 h^{-1}\,{\rm M_\odot}$ is injected at the halo center. The SMBHs accrete at the Bondi-Hoyle-Lyttleton rate, unless this rate is super-Eddington in which case the accretion rate is limited to Eddington. At both high and low accretion rates, a model for SMBH feedback is adopted; details are provided by \cite{Weinberger2017MNRAS}. \cite{Li2020ApJ} compared the SMBH populations in the TNG100 simulation to local SMBH-galaxy correlations. They found general agreement. However, even at a modest redshift of $z\sim3$, \cite{Natarajan2021arXiv210313932N} note that the implemented feedback prescriptions in the Illustris-TNG suite likely prematurely scuttle BH growth during phases of rapid accretion, which may explain why the simulations are unable to reproduce the observed high-redshift quasar population \citep[see also][]{Habouzit20222022MNRAS.511.3751H}.

Our work relies on the halo merger trees published by the IllustrisTNG collaboration, but not on the corresponding SMBH modeling. We adopt the Sublink galaxy merger trees, which are described in detail by \cite{Rodriguez2015MNRAS}. These merger trees generally agree with past theoretical and semi-empirical models of the galaxy merger rate, as well as observational constraints on the major-merger rate of medium sized ($M_*\ge 10^{10}\,M_\sun$ galaxies). There are discrepancies between the observed and predicted major-merger rate of massive ($M_*\ge 10^{11}\,M_\sun$ galaxies), but observations have not converged on the expected result.

We use the Sublink merger trees to construct a sub-tree consisting of each halo and at most two of its progenitors. When a halo has more than two progenitors, we adopt the two with the most massive histories behind them; i.e., those progenitors for which the sum of the masses of the galaxies in their merger histories are the largest. We choose to adopt this definition rather than simply taking the two most massive progenitors following the discussion in \cite{Rodriguez2015MNRAS}: adopting the most massive progenitors leads to arbitrariness when two progenitors have similar masses.

\subsection{Assigning SMBHs to galaxies} \label{sec:BH_gal}

\begin{figure*}
    \centering
    \includegraphics[width=0.99\textwidth]{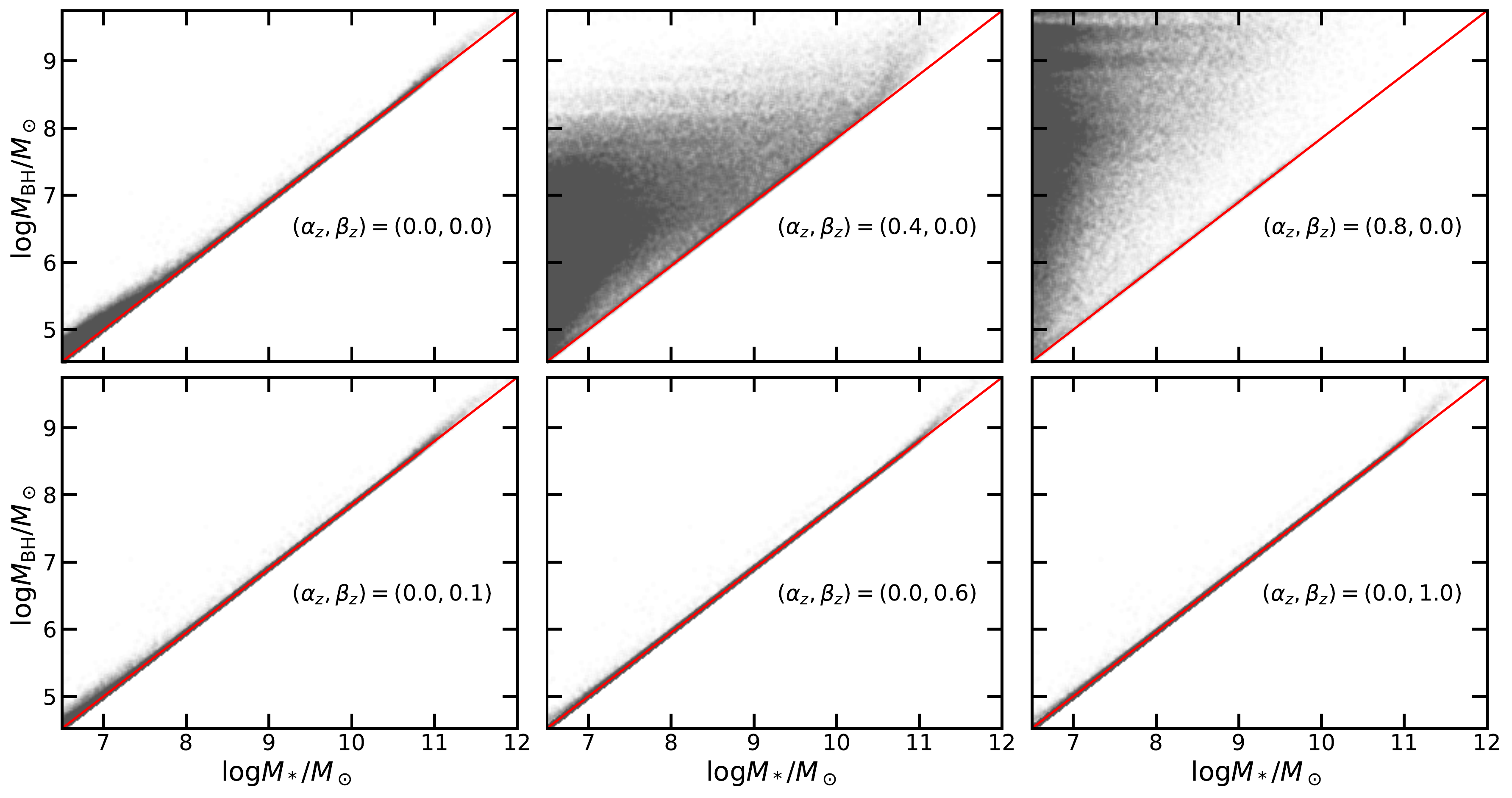}
    \caption{Example $z=0$ black hole populations produced by a range of $(\alpha_z, \beta_z)$ pairs. The top panels show the effects of varied $\alpha_z$ with fixed $\beta_z$ and the bottom panels show the effects of varied $\beta_z$ with fixed $\alpha_z$. In both cases, although the effects are more extreme for the range of $\alpha_z$ considered, large values of the redshift evolution parameters cause black holes that are unphysical: they significantly deviate from the fiducial local $M_{\rm BH}-M_*$ relation, shown in red.}
    \label{fig:fig1}
\end{figure*}

Although the TNG100 simulation includes SMBHs, we aim to constrain the range of feasible GWB amplitudes {\it without} making any assumptions about SMBH seeding or growth beyond what is definitively known from observations. There are two key observational constraints on SMBHs: in the local universe, well-constrained relations between SMBH mass and galaxy properties exist; in the local universe and at higher redshifts, the quasar luminosity function sets constraints on the number of accreting SMBH. In this section we focus on the former observation, in Section~\ref{sec:QLF}, we consider the quasar luminosity function.

We assign SMBHs to galaxies such that they are consistent with known SMBH-galaxy relations. SMBH masses are known to be correlated with galaxy central stellar velocity dispersion, stellar mass, S{\'e}rsic index, and star formation, among other galactic parameters. In this work, we focus on the SMBH mass -- stellar mass ($M_{\rm BH}-M_*$) correlation. We adopt this correlation because galaxy stellar masses are readily available from IllustrisTNG, whereas central (e.g., bulge) stellar velocity dispersions have not been calculated for the entire galaxy catalog for any of the TNG simulations.

The $M_{\rm BH}-M_*$ relation has been studied extensively in the local universe, while measurements of this correlation at high-redshift are complicated by the difficulty of obtaining accurate SMBH and stellar masses for distant galaxies. The local $M_{\rm BH}-M_*$ relation is often parameterized as
\begin{gather}
    \log \frac{M_{\rm BH}}{M_\odot} = \mathcal{N}\bigg(\alpha_0 + \beta_0 \log \frac{M_*}{10^{11}\,M_\odot}, \sigma_0\bigg), \label{eq:localMbhM*}
\end{gather}
Typical parameters are $\alpha_0 \approx 7.4-9$ and $\beta_0 \approx 0.4-1$, dependent on the population of galaxies considered (e.g., AGN, ellipticals; \citealp{Reines2015ApJ}). Comparing these observational results to simulations is complicated by differences in how stellar masses are measured observationally and in simulations, but \cite{Li2020ApJ...895..102L} showed that, after accounting for these effects, the default SMBH seeding and growth prescription in Illustris-TNG simulations produces populations with SMBH masses that are generally consistent with the observations. Hence, throughout this work, we will not compare directly the the observed best-fit SMBH mass -- stellar mass relation. Instead, we adopt the local $M_{\rm BH}-M_*$ relation measured in TNG100, using the SMBH populations included in that simulation. We fit Equation~\ref{eq:localMbhM*} to the stellar and SMBH masses all galaxies in the $z=0$ snapshot of the simulation and found $\alpha_0=8.545$ and $\beta_0=0.9496$, which are consistent with the observational results quoted above. We only considered galaxies with $\log M_*/M_\odot > 10$ in this fit to ensure that resolution effects will not change our result; we include the full galaxy population in the rest of our analysis.

To assign each galaxy an SMBH that satisfies the local $M_{\rm BH}-M_*$ relation, we adopt the following procedure. We assign initial SMBHs to galaxies using a generalized version of Equation~\ref{eq:localMbhM*} with linear redshift evolution:
\begin{gather}
    \log \frac{M_{\rm BH}}{M_\odot} = \mathcal{N}\bigg(\alpha(z) +\beta(z) \log \frac{M_*}{10^{11}\,M_\odot}, \sigma_0\bigg); \\\nonumber
        \alpha(z) = \alpha_z z + \alpha_0, \beta(z)  = \beta_z z + \beta_0.
\end{gather}
We assume a constant intrinsic scatter $\sigma_0 = 0.47$ dex for simplicity. We adopt the values of $\alpha_0$ and $\beta_0$ that we measured from the TNG100 simulation. $\alpha_z$ and $\beta_z$ are fit parameters that we will vary. Increasing $\alpha_z$ will cause the SMBH masses for redshifts $z>0$ to be higher than for the case $\alpha_z=0$; i.e., it will produce an overall more massive SMBH population at high redshift. Increasing $\beta_z$ causes the SMBH masses for galaxies with large stellar masses to be higher than for $\beta_z=0$, while the SMBH masses for galaxies with small stellar masses will be lower than for $\beta_z=0$.

\begin{figure*}
    \centering
    \includegraphics[width=0.99\textwidth]{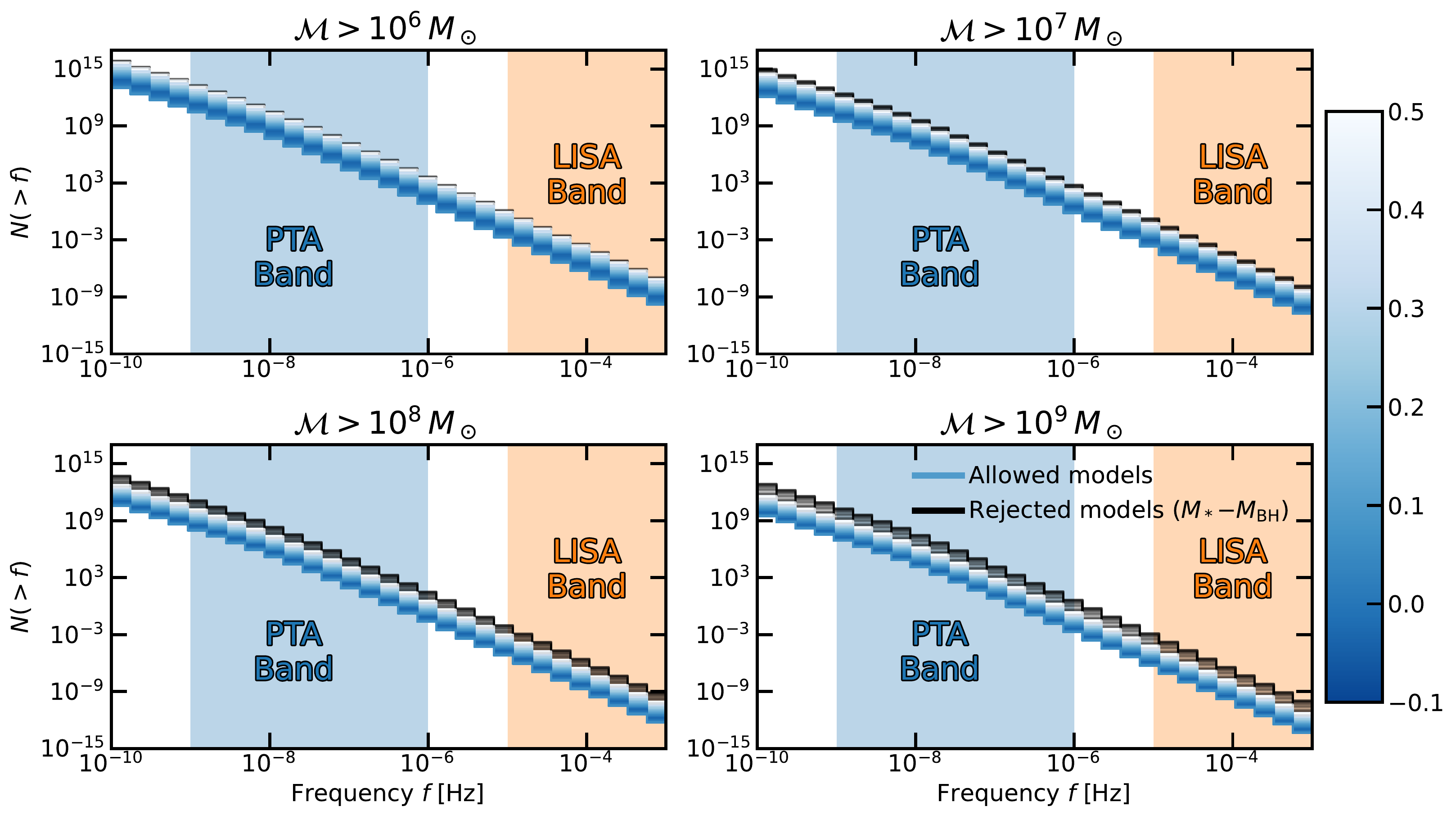}
    \caption{The number of binaries emitting as a function of logarithmic frequency. Models allowed by the local $M_{\rm BH}-M_*$ relation are shown in blue, where the shade of blue reflects the total amplitude of the GWB. Each panel shows the contribution from SMBHBs above a given chirp mass, as noted in the titles. The PTA band is highlighted in blue and the LISA band is highlighted in orange. }
    \label{fig:fig3}
\end{figure*}

This prescription will not necessarily produce a realistic SMBH population. First, we are implicitly assuming certain properties of the SMBH population that may not be realistic; i.e., that an $M_{\rm BH}-M_*$ relation exists at all redshifts. This assumption will not significantly change the results of this work. Our goal is to explore the amount by which we can modify the amplitude of the GWB by invoking modified redshift evolution, irrespective of the absolute value of the amplitude, and this model gives sufficient freedom for us to achieve this goal. However, because of this assumption, we primarily consider the change in the GWB amplitude $\Delta \log A_{\rm yr}$ relative to a fiducial value (arbitrarily picked to be the amplitude for $\alpha_z = \beta_z = 0$), rather than the absolute value $\log A_{\rm yr}$.

More importantly, for a certain range of $(\alpha_z, \beta_z)$, the SMBH mass for a fixed stellar mass will significantly decrease with time. This is unphysical, so we ensure that SMBH masses only grow with time using a simple prescription. Beginning at the highest redshifts, we identify each SMBH that has been assigned a mass smaller than the sum of those of its progenitors. We modify its mass to equal the sum of its parents' masses, and adjust each of its child SMBHs accordingly. This procedure will cause some SMBHs to fall off of the assumed $M_{\rm BH}-M_*$ relation. While there is no a-priori reason that the populations must satisfy the simple, linear $z\gg0$ $M_{\rm BH}-M_*$ relation, we must require consistency with the local SMBH mass -- stellar mass relation. Hence, we reject any model that has an SMBH more than $5\sigma$ deviant from the $M_{\rm BH}-M_*$ relation in the lowest redshift bin.

Following this prescription, for a given $\alpha_z$ and $\beta_z$, we are able to assign an SMBH to each galaxy in TNG100. We show example $z=0$ SMBH populations in Figure~\ref{fig:fig1}. We also flag any ($\alpha_z$, $\beta_z$) pairs that produce unrealistic SMBH populations (i.e., those for which the final SMBH population has an SMBH more than $5\sigma$ deviant from the $M_{\rm BH}-M_*$ relation in the lowest redshift bin).

\subsection{The amplitude of the GWB}

\begin{figure*}
    \centering
    \includegraphics[width=0.8\textwidth]{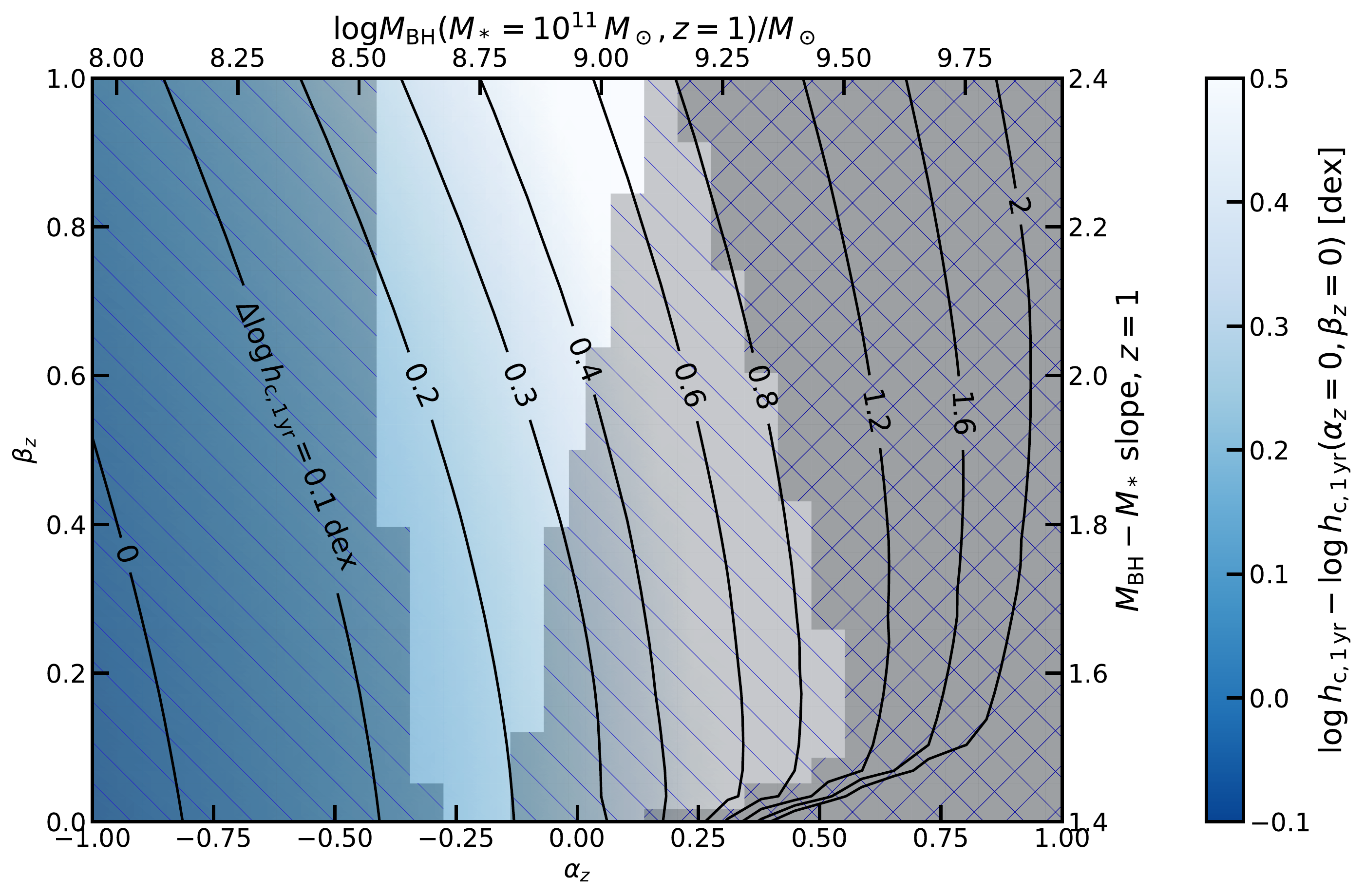}
    \caption{The GWB amplitude for varied values of $(\alpha_z, \beta_z)$, the parameters that control the redshift evolution of the $M_{\rm BH}-M_*$ relation. $\alpha_z$ is shown on the lower x-axis and $\beta_z$ is shown on the left y-axis. Equivalently, the SMBH mass for a $M_* = 10^{11}\,M_\odot$ galaxy at $z=1$ is shown on the upper x-axis and the slope of the $M_{\rm BH}-M_*$ relation at $z=1$ is shown on the right y-axis. The color and contours show the amplitude of the background relative to $(\alpha_z, \beta_z) = (0,0)$. The shaded region with hatches oriented towards the bottom-right corner shows the parameter space excluded by the QLF constraints and the shaded region with hatches oriented towards the top-right corner show the parameters space excluded by the local $M_{\rm BH}-M_*$ relation. Within the allowed subset of parameter space, we see variations in the background amplitude of ${\sim}0.5$ dex.}
    \label{fig:fig2}
\end{figure*}

We can now compute the GWB amplitude. First, we generate a population of SMBHs in merging galaxies for a given $\alpha_z$ and $\beta_z$. We assume that upon galaxy merger events bound SMBHBs are immediately formed at the critical separation where GW emission dominates the orbital hardening, such that the evolution of the orbital separation $a$ is given by
\begin{equation}
    \frac{d a_{\rm GW}}{dt} = -\frac{-64 G^3 M_1 M_2 (M_1+M_s)}{5 c^5 a^3},
\end{equation}
for a binary with masses $M_1$ and $M_2$. 

We assume circular binaries. Thus, following \cite{Sesana2008MNRAS}, we can calculate the GWB amplitude. We first consider the number of SMBHBs emitting per frequency and chirp mass bin. For a population of SMBHBs with comoving number density per unit chirp mass $\frac{dn}{d\Mch}$, this is given by
\begin{gather}
    \frac{d^2N}{d\Mch d \ln f} = \frac{dn}{d\Mch} \frac{dV_c}{dz} \frac{d z}{dt} \frac{dt}{d \ln f}\nonumber \\
    = \frac{dn}{d\Mch} \bigg( \frac{4 \pi d_L^2}{(1+z)^2} \bigg) \bigg( \frac{5}{96} \pi^{-8/3} \Mch^{-5/3} f_r^{-8/3}\bigg)  \\
    \frac{dN}{d \ln f}= f_r^{-8/3} \frac{1}{V_{\rm box}} \frac{20 \pi}{96} \pi^{-8/3} \sum_i  \frac{d_{L,i}^2 \Mch_i^{-5/3}}{(1+z_i)^2}. \label{eq:dNdf}
\end{gather}
$d_L$ denotes luminosity distance and $f_r$ rest-frame frequency. In the final step, we computed the total number of binaries emitting in logarithmic frequency bins for discrete binaries in a simulation box of size $V_{\rm box}$.

The characteristic strain for a population of SMBHBs evolving under GW emission alone is then given by
\begin{gather}\
    h_c^2(f {\mid} \alpha_z, \beta_z) = 
    \frac{4}{\pi f^2} \int_0^{\infty} dz \int_0^\infty d\mathcal{M} \frac{d^2n}{dz d\mathcal{M}} \frac{1}{1+z} \frac{d E_{\rm gw}(\mathcal{M})}{(d\ln f_r)}, \nonumber \\
    \frac{d E_{\rm gw}(\mathcal{M})}{(d\ln f_r)} = \frac{\pi^{2/3}}{3} \mathcal{M}^{5/3} f_r^{2/3}.
\end{gather}
Here, $f$ is the observed frequency of the GW signal and $f_r = (1+z)f$ is the rest-frame frequency for redshift $z$. The chirp mass is given by $\mathcal{M} = \mu^{3/5} M^{2/5}$ for reduced mass $\mu$ and total mass $M$. $\frac{d^2n}{dz d\mathcal{M}} = \frac{d^2n}{dz d\mathcal{M}}(\alpha_z, \beta_z)$ is the SMBHB number density per unit redshift and chirp mass. 

\begin{figure*}
    \centering
    \includegraphics[width=0.8\textwidth]{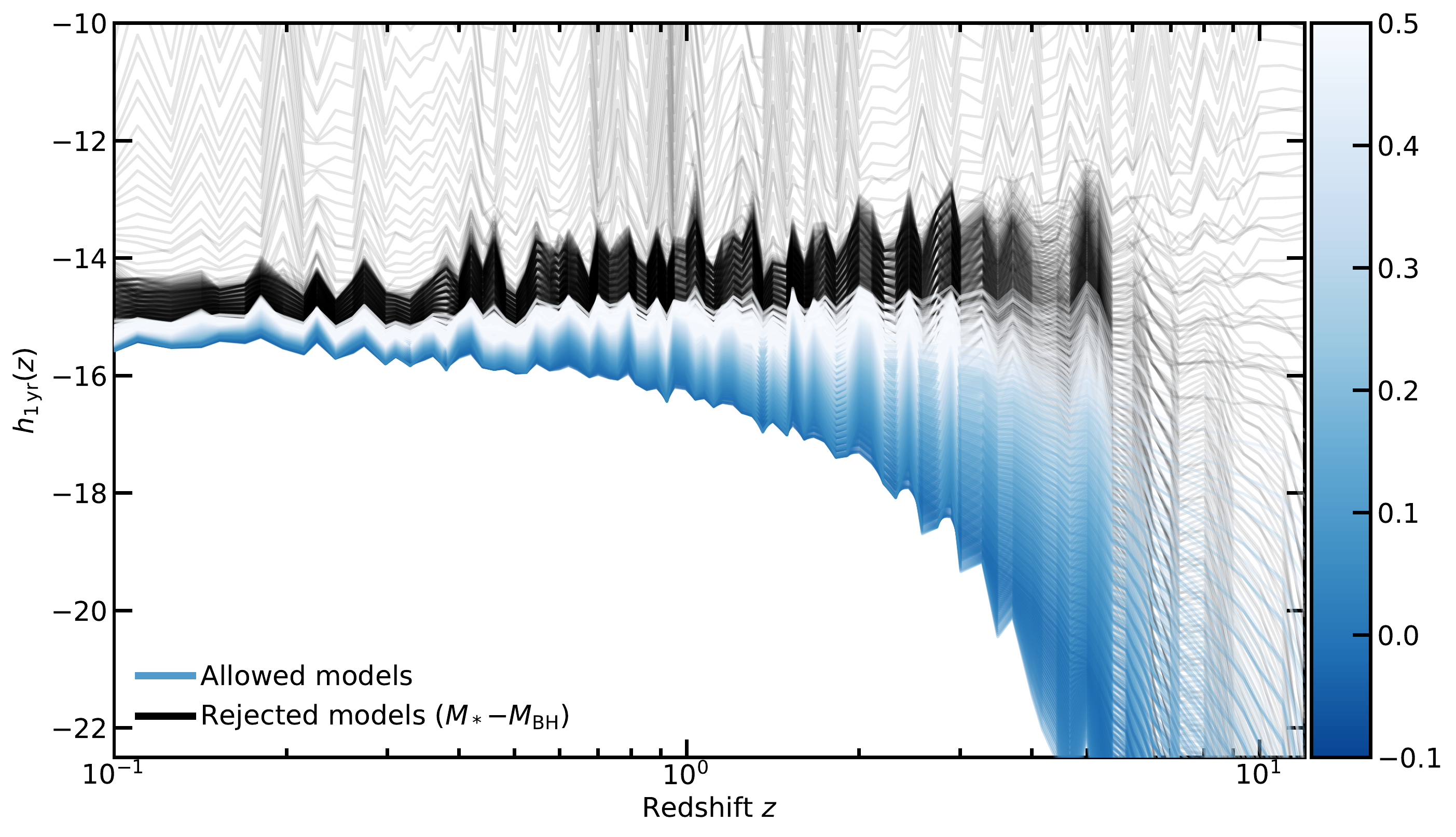}
    \caption{The contribution of each redshift to the total GWB amplitude. Models allowed by the local $M_{\rm BH}-M_*$ relation are shown in blue, where the shade of blue reflects the total amplitude of the  background. The $(\alpha_z, \beta_z) = (0,0)$ model is outline in dark blue. Black models are excluded by the local $M_{\rm BH}-M_*$ relation. We do not show constraints from the QLF in this Figure, but instead refer the reader to Figure~\ref{fig:fig2}. For models with higher background amplitudes, the local contribution to the background is slightly higher, but more significantly, the contribution of high redshifts is increased relative to that from low redshifts. }
    \label{fig:fig4}
\end{figure*}

For a discrete population of SMBHBs in a volume $V_{\rm box}$, this equation can be simplified to
\begin{equation}
    h_c^2(f {\mid} \alpha_z, \beta_z) = \frac{4\pi}{3 c^2} (2\pi f)^{-4/3} \sum_i \frac{1}{(1+z_i)^{1/3}} \frac{(G \mathcal{M}_i)^{5/3}}{V_{\rm box}},
\end{equation}
where $i$ indexes each SMBHB pair. We adopt this expression to calculate the GWB amplitude. 

\subsection{Ensuring consistency with the quasar luminosity function} \label{sec:QLF}

We calculate the GWB amplitude for a grid of models where we vary $\alpha_z$ and $\beta_z$. Not all of these models will be observationally feasible: they may not satisfy the local $M_{\rm BH}-M_*$ relation, and we flag these models as discussed in Section~\ref{sec:BH_gal}, or they may be inconsistent quasar luminosity function (QLF), which we discuss here. In brief, we require that the simulated number of luminous quasars be reasonably consistent with that predicted by the observed QLF. We do not expect to perfectly reproduce the QLF given our simplified SMBH prescription and that even the default SMBH formation model in TNG100 cannot reproduce the QLF, so we only require loose agreement.

In detail, we consider the QLF model from \cite{Shen2020MNRAS}. \cite{Shen2020MNRAS} compiled observations in the rest-frame IR, B band, UV, soft, and hard X-ray and combined them with an assumed quasar SED, bolometric correction model, and extinction model to derive an accurate QLF at redshifts from $z=0{-}7$. They fit the QLF to a flexible parameterization; we adopt the ``global fit A'' model. From this model, we can calculate $\phi_{\rm bol}(L_{\rm bol}, z) = \frac{dn}{d\log L_{\rm bol}}$, where $n = n(L_{\rm bol}, z)$ is the number density of quasars. However, we do not compare to the exact form of the QLF. Given our over-simplified assumption of a linearly evolving $M_*-M_{\rm BH}$ relation, we do not expect to be able to replicate the exact QLF. Instead, we simply aim to replicate the number of luminous SMBHs: $n(L_{\rm bol} > L_{\rm min.}, z) = \int_{L_{\rm min.}}^{\infty} \phi_{\rm bol}(L_{\rm bol}, z) d L_{\rm bol}$. We adopt $L_{\rm min.} = 10^{45}$ erg s$^{-1}$ and we have tested that increasing this value does not significantly reduce the allowed variation in the stochastic background amplitude. 

To compare to the observed $n(L_{\rm bol} > L_{\rm min.}, z)$, we must predict this value for each snapshot. We first calculate the difference in the mass of each SMBH between time steps, $\Delta M_{\rm BH}$. In the case of merging SMBHs, we take the difference between the sum of the parent masses and the child mass. We assume this mass change is due to accretion, and adopt an accretion rate equal to the average accretion rate between timesteps: assuming the change in look-back time between time steps is $\Delta t_{\rm lb}$, the accretion rate is $\dot{M}_{\rm BH} = \Delta M_{\rm BH}/\Delta t_{\rm lb}$. We convert the accretion rate to a luminosity as $L_{\rm bol} = \epsilon_r \dot{M}_{\rm BH} c^2$, where $\epsilon_r \sim 0.1$ is the radiative efficiency. The SMBH radiative efficiency is poorly constrained, so we include an uncertainty of a factor ${\sim}2$ by, for each SMBH in the simulation, drawing $\epsilon_r$ from a uniform distribution: $\epsilon_r \sim \mathcal{U}(0.05, 0.2)$. We then can calculate $n_{\rm pred.}(L_{\rm bol} > L_{\rm min.}, z)$ by counting the number of SMBHs with luminosities greater than $L_{\rm min.}$.

We define a consistent model as one for which $n_{\rm pred.}(L_{\rm bol} > L_{\rm min.}, z)$ is within $10\sigma$ of $n(L_{\rm bol} > L_{\rm min.}, z)$. Although this is does permit models which substantially disagree with the QLF, we choose such a loose $10\sigma$ constraint because, as discussed earlier, we do not expect to be able to reproduce the exact QLF. First, there are significant uncertainties in the observed QLF. In particular, the contribution from highly dust obscured AGN at high redshift is poorly constrained and observations of the cosmic X-ray background support the presence of a population of Compton-thick sources \citep{Comastri2015A&A...574L..10C}, so the number of quasars may be underpredicted. Moreover, our model is simplified, both in our SMBH seeding and evolution prescription and in our calculation of the quasar luminosities. While we should be able to produce quasars that somewhat reproduce the observe QLF, such a model cannot be expected to agree within a few $\sigma$ with observations. Even the more nuanced, physically motivated SMBH prescription included in the TNG simulations significantly overpredicts the QLF for $z\ge 3$ at the high luminosities \citep[][]{Weinberger2018MNRAS.479.4056W}. The agreement at lower redshift is better, although there are disagreements at the low-luminosity end. Moreover, as we will show in the next section, our requirement that the simulations agree with the $M_{\rm BH}-M_*$ relation, regardless of the QLF, sets a strong constraint on feasible models with high GWB amplitudes. Adding the QLF constraints does not significantly change the maximum possible amplitude of the GWB.

\section{Results} \label{sec:results}

Following the prescription described in Section~\ref{sec:methods}, we calculate the amplitude of the GWB and the number of luminous quasars for a grid of $(\alpha_z, \beta_z)$. We consider the following parameter ranges:
\begin{gather}
    \alpha_z \in [-1,1] \nonumber \\
    \beta_z \in [0,1].
\end{gather}
These parameter ranges preferentially sample pairs of $(\alpha_z, \beta_z)$ that will produce a higher $A_{\rm 1\,yr}$, as this is the regime that we are most interested in studying. We generate SMBH populations for a uniform grid in this parameter space with dimensions ${\sim}30 \times 30$. 

In Figure~\ref{fig:fig3}, we show the number of SMBHBs emitting as a function of frequency for different $(\alpha_z, \beta_z)$, calculated with Equation~\ref{eq:dNdf}. We have noted the approximate range of frequencies that PTAs and the Laser Interferometer Space Antenna \citep[LISA;][]{LISA2023LRR....26....2A} are sensitive to. These results are consistent with, e.g., results from \citet{Sesana2008MNRAS} and \citet{Ravi2015MNRAS.447.2772R} (their Figures 5 and 4 respectively). 

The GWB amplitude as a function of $\alpha_z$ and  $\beta_z$ is shown in Figure~\ref{fig:fig2}. Rather than show the absolute value of the GWB amplitude, we show the deviation relative to $(\alpha_z, \beta_z) = (0,0)$, although this case is excluded by the QLF because it over-produces high luminosity quasars at redshift $z\sim3$. The absolute value of the GWB amplitude depends significantly on, e.g., our choice for the local $M_{\rm BH}-M_*$ relation, and thus is subject to uncertainty. Our assumption of a perfectly linear $M_{\rm BH}-M_*$ relation with constant scatter tends to result in GWB amplitudes that are higher than those predicted by the default SMBH relation in Illustris (${\sim}10^{-15}$), except for the lowest values of $(\alpha_z, \beta_z)$. Instead, by looking at the level of variation feasible by changing the high redshift population, we can gain a sense for how much one might be able to change the amplitude relative to any given local SMBH constraints. We verified that if we use the SMBHs simulated in TNG100 we essentially reproduce the results of \citet{Kelley2017MNRAS}, although they used an earlier version of the Illustris simulation. 

The shaded and hatched regions in Figure~\ref{fig:fig2} indicate the parameter space that is excluded by the local $M_{\rm BH}-M_*$ relation and the QLF respectively. The local $M_{\rm BH}-M_*$ relation excludes the region with $\alpha_z \gtrsim 0.5$, while the QLF constrains both the low $\alpha_z$ region, which contains too few high-redshift luminous quasars, and the high $(\alpha_z, \beta_z)$ region, which overproduces the high-redshift quasar population. 

Within the allowed parameter space, the GWB amplitude varies by ${\sim}0.3$ dex. Figure~\ref{fig:fig4} shows the contribution of each redshift to the final amplitude. While the local contribution does change slightly due to slightly higher mass SMBHs, the slope of the amplitude contributions with redshift becomes significantly shallower for those SMBH populations with higher total amplitudes. In other words, the contribution of high redshift objects is more significant for the higher amplitude models. These high amplitude models are those with the highest allowed values of $(\alpha_z,\beta_z)$; i.e., these are the models that establish a larger population of high mass SMBHs at $z\gg0$ relative to the models with low $(\alpha_z,\beta_z)$. In these models, the high mass SMBHs are first established at high redshifts and then do not evolve significantly. In the low amplitude models, the SMBH masses are lower at high redshift, and they slowly increase with cosmic time.

\section{Summary and discussion} \label{sec:imp}

In summary, we have used a simple model for SMBH seeding and growth to predict the possible range of amplitudes of the GWB from SMBHBs in the PTA band, subject to basic observational constraints. We assumed a fiducial model for the merging galaxy population from TNG100, and assigned SMBHs to galaxies ourselves. We found that, if SMBHs form most of their mass at high redshifts, the GWB amplitude can be increased by a few tenths of a dex (see Figure~\ref{fig:fig2} for a summary of constraints). This increase is primarily due a large contribution to the GWB from high redshifts relative to that from the local universe, in contrast with earlier models for the GWB (Figure~\ref{fig:fig4}). This can be intuited as arising from the poor constraints on high-redshift QLF, and the difficulty that existing simulations have in reproducing even the known population of high-redshift quasars, together with the ``negative $K$-correction'' of the GW signals from individual SMBHBs. Our results imply that the surprising indications of a louder than expected GWB in recent PTA data \citep{Arzoumanian2020ApJ...905L..34A,Goncharov20212021ApJ...917L..19G,Chen20212021MNRAS.508.4970C} may be explained through a heretofore unmodeled population of high-redshift SMBHBs. In this scenario, the SMBH population must form most of its mass density by $z\sim1$.

Our work has several caveats and extensions. First, our conclusions will be somewhat modified if  stricter consistency was specified between the observed and predicted QLFs. This in turn would require a physical model for accretion and radiation, rather than the simple assumption of self-consistent, steady growth with a fixed radiative efficiency that we make. We anticipate however that, for the near future, the local SMBH mass -- stellar mass relation will remain more constraining. Second, more complex forms than a simple power-law for both the SMBH mass --- stellar mass relation and the redshift-evolution in that relation are possible, and remain to be explored. Likewise, we could include constraints from other SMBH -- host galaxy relations; e.g., we could have adopted the SMBH -- galaxy bulge mass relation. Third, we have not accounted for various physical mechanisms that may affect the formation and evolution of SMBHBs, including the possibilities of delayed merging \citep[e.g.,][]{Tremmel2018MNRAS.475.4967T} and triple interactions \citep[e.g.,][]{Volonteri2003ApJ...582..559V}, ejected SMBHs post-merger \citep[e.g.,][]{Ricarte2021ApJ...916L..18R}, and environmental torques on SMBHBs \citep[e.g.,][]{Ravi2014MNRAS.442...56R}. These effects are all more likely to be significant in the early universe, and thus need to be accounted for in more detailed modeling of the high-redshift contribution to the GWB from SMBHBs. Fourth, we do not assume any physical model for SMBH seeds \citep{Latif2016PASA...33...51L, Natarajan2017ApJ...838..117N}, but instead assume that SMBHs simply grow along with their host galaxies (at sub-Eddington rates) according to various SMBH-galaxy scaling relations. Our scenario is roughly consistent with models for accretion driven growth of SMBHs towards the observed massive high-redshift quasars \citep[e.g.,][]{Tanaka2009ApJ...696.1798T}. However, it is possible that physically motivated seeding and accretion prescriptions may significantly alter the route SMBHs take towards the SMBH-galaxy scaling relations \citep[e.g.,][]{Bonoli2014MNRAS.437.1576B,Ricarte2018MNRAS.474.1995R,Ricarte2018MNRAS.481.3278R}.

The prospect of a stronger than expected GWB amplitude arising from a population of SMBHBs in the early universe can be tested with current and upcoming instruments. If a significant portion of the GWB amplitude is produced by high-$z$ SMBHBs, then we might expect fewer local, continuous wave sources to be detectable \citep[][]{Arzoumanian2023arXiv230103608A}. In addition to such constraints from PTAs, observations from instruments including the James Webb Space Telescope (JWST) and the Laser Interferometer Space Antenna (LISA) will be extremely valuable. JWST will reveal a far more extensive population of SMBHs at high redshifts than is currently known \citep{Jeon2023arXiv230407369J,Harikane2023arXiv230311946H, Koceviski2023arXiv230200012K,Furtak2022arXiv221210531F,Larson2023arXiv230308918L,Bogdan2023arXiv230515458B,Maiolino2023arXiv230512492M}, enabling detailed measurements of the high-redshift $M_{\rm BH}-M_*$ relation, among other SMBH-galaxy connections. LISA will also provide key constraints on the SMBHB population and merger rate at high redshifts \citep{LISA2023LRR....26....2A}. It will be able to detect lower mass $M_{\rm BH}\sim 10^{4-7}\,M_\odot$ binaries at higher redshifts $z\gtrsim 1$ and, possibly, individual, bright sources at higher black hole masses $M\gtrsim 10^8\,M_\odot$. The strength of both signals will depend significantly on the mass distribution of the high-$z$ SMBHB population. As shown in Figure~\ref{fig:fig3}, the number of sources in the LISA band depends sensitively on the adopted model for the SMBH-galaxy connection, which in turn determines the amplitude of the GWB in the PTA band, although there is significant uncertainty in the LISA signal due to unknown SMBH seeding physics and merger dynamics \citep{Ricarte2018MNRAS.481.3278R, Ricarte2018MNRAS.474.1995R}. LISA measurements of the coalescing SMBHB population will provide valuable data on both the local and high redshift SMBHB demographics.

\begin{acknowledgments}
We would like to thank Fabian Walter, Paul Lasky, and Ryan Shannon for useful discussions. We would like to thank Joseph Lazio, Priya Natarajan, and Jenny Greene for valuable feedback. This material is based upon work supported by the National Science Foundation Graduate Research Fellowship under Grant No. DGE‐1745301.
\end{acknowledgments}

\bibliography{sample631}{}
\bibliographystyle{aasjournal}

\end{document}